\documentclass[particles,article,accept,pdftex,moreauthors]{Definitions/mdpi} 

\newcommand{\be}{\begin{equation}}
\newcommand{\ee}{\end{equation}}
\newcommand{\beal}{\begin{aligned}}
\newcommand{\eeal}{\end{aligned}}
\newcommand\bea{\begin{eqnarray}}
\newcommand\eea{\end{eqnarray}}
\newcommand{\bec}{\begin{cases}}
\newcommand{\eec}{\end{cases}}
\newcommand{\tila}{{\tilde\alpha} }
\newcommand{\bsig}{{\bar\sigma} }

\firstpage{144} 
\pubvolume{7}
\issuenum{1}
\articlenumber{8}
\pubyear{2024}
\copyrightyear{2024}
\externaleditor{Academic Editor:Armen Sedrakian }
\datereceived{29 November 2023} 
\daterevised{9 February 2024} 
\dateaccepted{12 February 2024} 
\datepublished{16 February 2024 } 
\hreflink{https://doi.org/10.3390/ \linebreak particles7010008} 

\makeatletter
\let\c@lofdepth\relax
\let\c@lotdepth\relax
\makeatother
\usepackage{subfigure}

 \Title{Testing Higher Derivative Gravity Through Tunnelling}

\TitleCitation{Testing Higher Derivative Gravity Through Tunnelling}

\Author{Ruth 
 Gregory $^{1,2}$\orcidA{} and Shi-Qian Hu $^{1,}$*\orcidB{}}

\AuthorNames{Ruth Gregory, Shi-Qian Hu}

\AuthorCitation{Gregory, R.; Hu, S.-Q.}

\address{%
$^{1}$ \quad Department of Physics, King's College London,
The Strand, London WC2R 2LS, UK; ruth.gregory@kcl.ac.uk\\
$^{2}$ \quad Perimeter Institute, 31 Caroline Street North, Waterloo, 
ON N2L 2Y5, Canada}
\corres{Correspondence: shiqian.hu@kcl.ac.uk
}

\abstract{
Higher derivative terms in the gravitational action are natural from the perspective of quantum gravity, but are perceived as leading to a lack of well-posedness. The Gauss--Bonnet term has second-order equations of motion, but does not impact gravitational dynamics in 4D, so one might expect that it is not physically relevant. We discuss how signatures can show up in tunnelling processes and whether these will likely be physically accessible in Higgs vacuum decay.
}

\begin{document}
\section{Brief Overview of Seeded Vacuum~Decay}
\label{sec: Intro} 
One of the mysteries at the core of the conflict between gravity and particle physics is the nature of the vacuum. Assumptions about the vacuum underlie Hawking radiation and inflation. But,~how do we define the vacuum? The simplest way is to say it is the state of ``lowest energy''; however, this presupposes that there is a lower bound and that we live in it. 
This research arose from re-examining the methods behind the calculations of first-order phase transitions in the universe. In~nature, first-order phase transitions are usually triggered by impurities, but~the classic field theory description of a first-order transition, quantum tunnelling, is more or less the opposite---extremely symmetric and idealised. How do we translate this concept of impurity to a gravitational setting, and~what pitfalls arise? Further, if~we go beyond Einstein's theory, how does this modify the~results?

\textcolor{black}{Theories of modified gravity have been developed to address some of the issues that exist in General Relativity (GR), such as late time acceleration or~as an alternative to dark matter (see~\cite{Clifton:2011jh}  for a review). 
More recently, considering modifications to GR in the strong gravitational regime provides a more-robust discriminator when testing gravity near black holes observationally.} 
Among the alternatives to GR, higher curvature theories have a uniquely privileged position as they require no extra fundamental fields and can be motivated within a quantum gravity setting. 
While adding an arbitrary higher derivative term to the gravitational action can result in quartic (or higher) time derivatives, which leads to a lack of well-posedness, the~Gauss--Bonnet (GB) \cite{Lanczos:1938sf, Zumino:1985dp, Zwiebach:1985uq} term (the lowest-order Lovelock contribution~\cite{Lovelock:1971yv}) is quadratic in curvature while retaining only second-order field equations. 
\textcolor{black}{A further feature of relevance to this work is the fact that a general higher curvature term will not have a Birkhoff-type theorem for the uniqueness of black hole solutions. 
When one has a Birkhoff theorem, the~uniqueness of a black hole geometry is guaranteed. Thus, if~one is performing a path integral, the~geometries satisfying some given boundary constraints are now specified precisely, and~one can be sure that the saddle point has been fully identified. 
In the absence of this uniqueness theorem, the~possibility arises of multiple geometries having the same asymptotic form; hence, a background subtraction may not be unique, or~even defined. Therefore, for~the purpose of testing how higher curvature terms alter gravitational phenomenology in the context of vacuum decay, we focus on the GB term as an exemplar of a well-posed theory that has a Birkhoff uniqueness property~\cite{Charmousis:2002rc}.}

The reason for being interested in vacuum decay lies in the discovery of the Higgs boson~\cite{ATLAS:2012yve, CMS:2012qbp}. The~measurement of Standard Model (SM) parameters, particularly the Higgs and top quark masses, reveals the possibility of a lower vacuum energy in the effective potential of the SM Higgs field at high Higgs values~\cite{ATLAS:2013dos, CMS:2013fjq, ATLAS:2015yey, CMS:2015lbj}.
This suggests that our universe could be in a metastable vacuum that can undergo quantum tunnelling from our local minimum (false vacuum) to a global minimum (true vacuum) or exit point at a very different Higgs value. 
This decay is traditionally described by a process of bubble nucleation, where a bubble of true vacuum nucleates in the ambient false vacuum and, then, expands.  This process was described by Coleman and collaborators~\cite{Coleman:1977py, Callan:1977pt,Coleman:1980aw} in a sequence of papers computing the tunnelling rate via so-called bounce solutions: regular solutions of the Euclidean field equations, which can, then, be readily extended to a gravitational setting using the Gibbons--Hawking~\cite{Gibbons:1976ue} and Israel~\cite{Israel:1966rt} formalisms in~GR. 

An alternate decay process is via the Hawking--Moss (HM) \cite{Hawking:1981fz} instanton---a more-quantum-cosmology-motivated picture in which the universe instantaneously ``jumps'' to a higher vacuum energy corresponding to the (maximum) turning point of the potential, the~idea being that it will then evolve cosmologically to lower, or~true vacuum,\mbox{ energy~values. }

Both processes involve a Euclidean quantum gravity approach, in~which the instanton is a Euclidean solution, and~the leading order contribution to the probability of decay is the exponential of the tunnelling action---the difference between the initial false vacuum state and either the bounce action in the case of bubble nucleation or the higher 
\textcolor{black}{vacuum energy universe action in the case of HM. 
The main difference between HM and Coleman instantons is how one interprets the transition, particularly when considering the transition to a real-time Lorentzian interpretation. 
The bubble nucleation picture has a natural analytic continuation back to Lorentzian (real) time, and~trajectories found in the construction of the instanton rotate back to real time in a straightforward manner. 
The conventional interpretation is that one takes a constant $\tau$ slice of the instanton that has vanishing extrinsic curvature---i.e.,\ is ``static''---and, then, simply takes the analytically continued solution, $\tau\to -it$, 
 with the same initial conditions at $\tau=i t=0$. 
The corresponding real time trajectories can be found, for~example, in~\cite{Burda:2016mou, Brown:2007sd}.
The round Euclidean bubble transforms into a hyperbolic expanding Lorentzian bubble, which rapidly expands into the false vacuum.
The HM transition, however, is more fundamentally quantum. 
There is no analytic continuation and~no ``initial data surface'', as the whole universe has literally jumped into a new vacuum state; rather, the~new (unstable) local stationary point itself represents a new configuration of the universe that then rolls down into a new region of parameter space in a manner similar to (at least initially) the rolling of an inflationary~universe.}

\textcolor{black}{In both the HM and bubble nucleation} cases, the~original computations assume the maximal symmetry of the solutions: 
the bubble nucleates in a constant curvature background and preserves $SO(D)$ symmetry 
(in D-dimensions); the HM instanton is simply the difference in action between 
de-Sitter (dS) solutions with different $\Lambda$s. Given that first-order phase transitions in nature 
are typically initiated by impurities, it is clear that we need to make this picture more realistic by 
breaking some of these symmetries. While the general problem would require a full numerical 
calculation in GR, adding a black hole breaks a little of the symmetry while retaining most of 
the calculational control, allowing a more-general instanton to be constructed and the action 
computed~\cite{Gregory:2013hja,Burda:2015yfa,Burda:2015isa,Burda:2016mou}. 

Before proceeding with a summary of this picture, it is worth noting that, while quantum 
mechanical tunnelling is well tested and experimentally verified, quantum \emph{field theory} 
tunnelling is not. As~such, the~approximation of taking only the suppressed exponent and the 
semi-classical saddle point (on which gravitational calculations rely) is based on parallels with 
other field theory computations, and~the Euclidean approach for bubble nucleation has been 
challenged~\cite{Braden:2018tky}. For~the purposes of this presentation, we note these criticisms, 
but view analytic continuation of the ``time'' co-ordinate as a tool---to be used when helpful 
and~appropriate.

\subsection{Tunnelling \`a la~Coleman}

\textcolor{black}{In this subsection, we review the classic picture developed by Coleman and others describing vacuum decay} 
via the mathematical tool of analytically continuing to imaginary (Euclidean) time. Coleman's 
intuitive picture was to view the bubble (approximately) as a spherical bubble of true vacuum 
inside the false vacuum, where the bubble wall has an energy density equal to its tension. 
The energetics of the bubble, then, are a gain in energy from the interior being in a true vacuum, 
thus at lower energy than the false one, countered by the energy cost of the thin wall. 
We are, therefore, balancing area and volume. We can then find the instanton via a ``Goldilocks'' 
argument. Bubbles will fluctuate into existence, but~if they are too small, the large surface-area-
to-volume ratio causes them to re-collapse. On~the other hand, while a very large bubble has a 
better area-to-volume ratio, the~overall cost, or~action, of~creating it lowers the probability. 
There is, therefore, a ``just right'' size, where the action has an (unstable) stationary point---
the Goldilocks bubble---which is the instanton. In~terms of a formula, the energy differential is the cost of the wall less the gain in falling into true vacuum \textcolor{black}{(a more-detailed derivation can be found in~\cite{Coleman:1977py} and~a heuristic discussion also including gravity in~\cite{Gregory:2023eos})}:
\be
\delta E = 2\pi^2 R^3 \sigma - \frac{\pi^2}{2} \varepsilon R^4
\label{goldilocksE}
\ee
where $\sigma$ is the tension of the wall and $\varepsilon$ is the difference in energy between the 
true and false vacuum, assumed small.
This energy shift is stationary at $R=3\sigma/\varepsilon$, which corresponds beautifully to a 
calculation of the full Euclidean field equations. 
\textcolor{black}{Substituting this value of $R$ at the stationary point of \eqref{goldilocksE} gives the bounce action as} 
\be
{\cal B} = \frac{\pi^2 R^3}{2} (4\sigma - \varepsilon R) = \frac{27\pi^2 \sigma^4}{2\varepsilon^3}
\label{coleman1}
\ee
which then gives the leading-order, or~saddle point, contribution to the amplitude for tunnelling
${\cal P} \sim e^{-{\cal B}}$.

Of course, this is not the full story: energy gravitates; therefore, we should include the impact of
an infinite volume of false vacuum. This was worked out in the paper of Coleman and de Luccia 
(CDL) \cite{Coleman:1980aw} using the thin wall approximation described above. As~before, the~
instanton is a solution of the Euclidean Einstein equations with a spherical bubble separating true and
false vacuum solutions. Computing quantum processes with gravity is always delicate, as we do not have a consensus on 
a working theory of quantum gravity. Here, we will be taking the partition function approach of Gibbons
and Hawking~\cite{Gibbons:1976ue} and work within the saddle point approximation. It is worth noting that this focus
on the gravitational side of the problem does not address questions such as thermal corrections to the
scalar (Higgs) potential.

The gravitational action with a thin wall is 
\be
S_{E} = - \frac1{16\pi G} \int_{{\cal M}_+ \cup{\cal M}_-} d^4 x \sqrt{g} (R-2 \Lambda) 
+ \frac1{8\pi G} \int_{\partial{\cal M}_+ \cup \partial{\cal M}_-} d^3 x \sqrt{h} K
+ \int_{\cal W} d^3 x \sqrt{h} \sigma
\label{GibHawk}
\ee
and the CDL procedure is to construct a geometry that interpolates between the true and false vacua
across the wall, using the \textcolor{black}{Israel junction} conditions to calculate the wall trajectory. To~illustrate briefly, consider
tunnelling from positive (false) to zero (true) vacuum energy. A~Euclidean de-Sitter space has the 
geometry of a four-sphere with radius $\ell = \sqrt{3/8\pi G\varepsilon}$, and~zero vacuum energy is 
just flat Euclidean $\mathbb{R}^4$:
\be
ds^2 = 
\begin{cases}
d\rho^2 +  \ell^2 \sin^2 (\rho/\ell) d\Omega_{I\!I\!I}^2 & \text{false (de-Sitter)}\\
dr^2 + r^2 d\Omega_{I\!I\!I}^2  & \text{true}
\end{cases}
\ee

\textls[-15]{The 
 polar coordinates are chosen to centre on the wall, which sits at
$r_0 = \ell \sin(\rho_0/\ell) = R$}, which has an induced metric:
\be
h_{ab} = g_{ab} - n_a n_b
\ee
where the normal $n_a$ is proportional to $d\rho$ or $dr$. The~energy momentum of the wall
is $\sigma h_{ab}$, which is related to the jump in extrinsic curvature via the \textcolor{black}{Israel junction conditions:}
\be
K_{ab}^+ - K_{ab}^- = -\frac1R \left ( 1 - \sqrt{1 - \frac{R^2}{\ell^2} }
\right) h_{ab} = - 4 \pi G \sigma h_{ab}
\ee

Writing $\bar{\sigma} = 2\pi G\sigma$, this is solved by
\be
R_0 = \frac{4\bar{\sigma}\ell^2}{1 + 4 \bar{\sigma}^2 \ell^2}
\label{RCDL}
\ee
which 
 can, then, be substituted into the Euclidean action to obtain
\be
{\cal B}_{CDL} = S_E(\text{instanton}) - S_E(dS)
=\frac{\pi \ell^2}{G} \frac{16 \bar{\sigma}^4 \ell^4}
{(1 + 4 \bar{\sigma}^2 \ell^2)^2}
\label{BCDL}
\ee

Noting the expressions for $\ell$ and $\bsig$, we see that this expression
\be
{\cal B}_{CDL} = \frac{27 \pi^2 \sigma^4}{2 \varepsilon^3 (1 + 6\pi G \sigma^2)^2}
\ee
has the correct $G\to0$ limit \eqref{coleman1}.

\subsection{Seeded~Tunnelling}

The description of tunnelling given above is extremely idealised with maximal symmetry and no 
inhomogeneity; however, most first-order phase transitions are triggered by impurities in the
system. Taking this idea on board, Refs.~\cite{Gregory:2013hja,Burda:2015yfa,Burda:2015isa,Burda:2016mou}
considered a gravitational impurity, in~the guise of a black hole, that breaks the maximal symmetry, but~
retains enough symmetry to allow analytic analysis in the case of the thin wall. \textcolor{black}{Black-hole-seeded tunnelling in Einstein gravity is recapped in this subsection to set the stage for the inclusion of the GB term. We show the comparison rate between Hawking evaporation and tunnelling, as~well as the HM instanton. }

The instanton geometry now becomes that of a thin wall solution interpolating between two
different Schwarzschild black holes with different cosmological constants. The~problem of finding the
geometry of a wall in the presence of a bulk black hole reduces to a Birkhoff theorem,
which was first generalised to arbitrary $\Lambda$ and black hole curvature in~\cite{Bowcock:2000cq}
\textcolor{black}{(and arbitrary dimension in~\cite{Gregory:2001xu})},
where a domain wall was also included in the analysis.
The integrability of the vacuum Einstein equations shows that the general solution is comprised of two exact
Schwarzschild (A/dS) geometries separated by a domain wall, which follows a possibly dynamical
trajectory $R(\lambda)$ centred with respect to the \mbox{black holes}.
\be
ds^2 = 
\begin{cases}
f_+(r) d\tau_+^2 + \frac{dr^2}{f_+(r)} + r^2 d\Omega_{I\!I}^2 & R>R(\lambda) \\
f_-(r) d\tau_-^2 + \frac{dr^2}{f_-(r)} +  r^2 d\Omega_{I\!I}^2 & R<R(\lambda)
\end{cases}
\label{bubblegeom}
\ee
with 
\be
f_\pm(r) = 1 - \frac{2M_\pm}{r} - \frac{\Lambda_\pm}{3} r^2
\ee

It is straightforward to find solutions to the \textcolor{black}{Israel junction conditions}, and~typically, for each seed mass $M_+$,
there is an allowed range for $M_-$ for the given vacuum energy differential. 
Although the methodology
for computing the action is slightly distinct depending on the sign of $\Lambda$, the~final result for the 
instanton action is relatively simple \textcolor{black}{\cite{Gregory:2013hja}}:
\begin{equation}
{\cal B}= {{\cal A}_+\over 4G}-{{\cal A}_-\over 4G} +{1\over 4G}
\oint d\lambda \left\{\left ( 2Rf_+ - R^2 f_+'\right)\dot{\tau}_+
- \left (2Rf_- -R^2 f_-'\right) \dot{\tau}_-\right\}
\label{bounceaction}
\end{equation}

It is, then, easy to see that, for each $M_+$, there is a unique $M_-$ that minimises this action, 
hence dominates the tunnelling amplitude. In~\cite{Gregory:2013hja}, it was shown that, for small $M_+$,
this minimal $M_-$ vanishes; hence, these instantons are perturbed CDL bubbles, whereas, 
for larger $M_+$, there is a remnant black hole, with~a mass determined by $M_+, \Lambda_+$ and 
$\Lambda_-$, this latter instanton being independent of Euclidean time. As~a static solution, the~action
of these instantons is particularly simple and~is just the entropy change in the geometry:
\be
{\cal B}_s = S_+-S_-
\ee
\textls[-15]{which corresponds neatly to a Boltzmann suppression for a tunnelling that \mbox{decreases entropy}.}

In~\cite{Burda:2015yfa,Burda:2015isa,Burda:2016mou}, the thick wall versions of these instantons were
constructed numerically to explore the impact on Higgs vacuum tunnelling, with~the result that the
black hole could potentially drastically increase the probability of vacuum decay. 
In the case of a black-hole-seeded phase transition, the~key comparison is with the rate of evaporation 
of the black hole, which is proportional to $1/M^3$, thus might be expected to be dominant at all
scales; however, the~prefactor of this power law, computed by Page~\cite{Page:1976df}, is suppressed,
so the question is more subtle. In~fact, when computing the area difference between seed and remnant 
black holes, the~mass screened by the thick Higgs bubble is significantly less than the mass of the
seed black hole, so the entropy difference is, to~a good approximation,
\be
{\cal B} = \pi (r_s^2-r_r^2) \sim 4\pi (M_s+M_r)(M_s-M_r) = 8\pi M_s \delta M
\ee

\begin{figure}[H]
\includegraphics[width=5in]{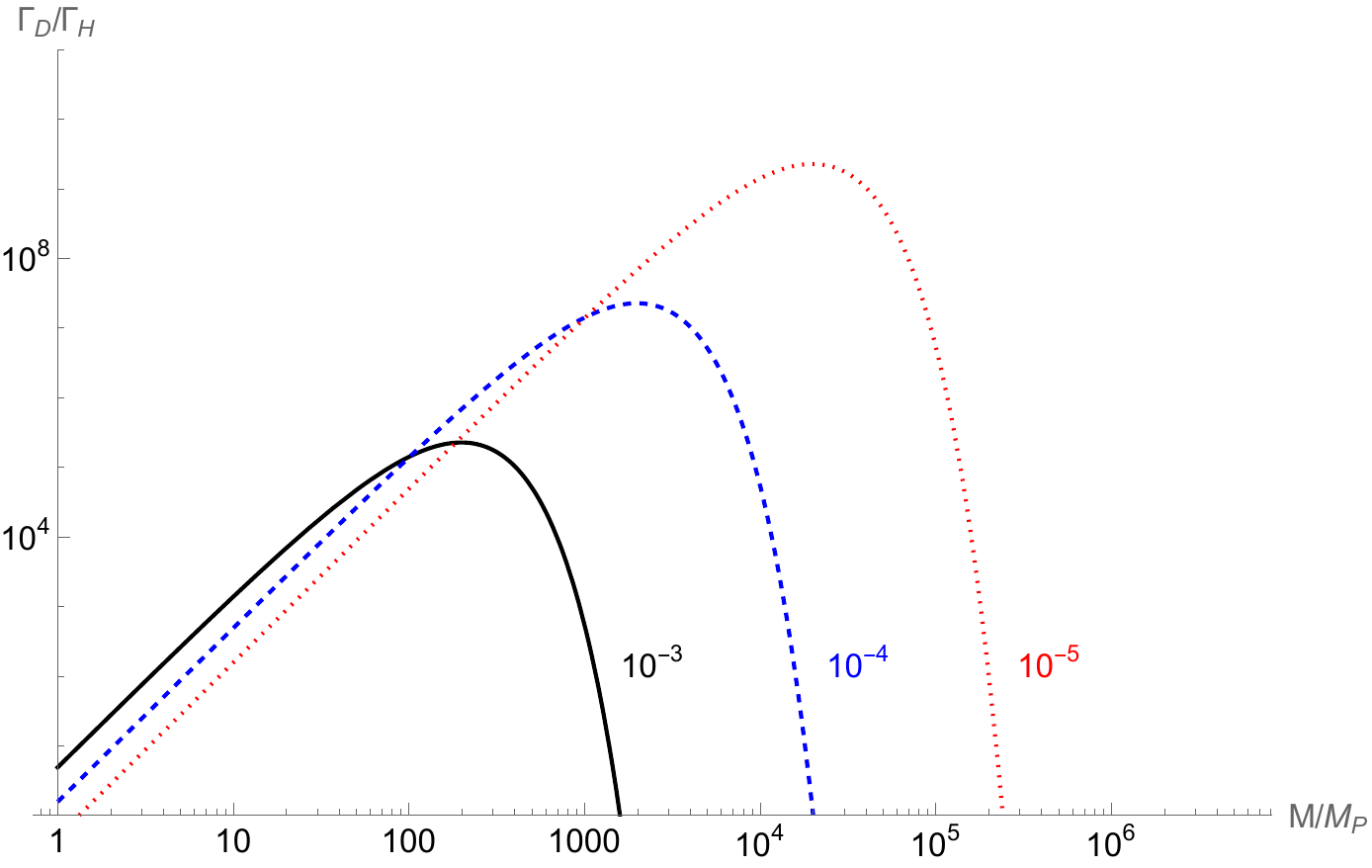}
\caption{The 
 approximation for $\Gamma_D/\Gamma_H$ as given in \eqref{approxratio} as a function of seed mass for the 
differential seed ratio masses indicated.}
\label{fig:branch}
\end{figure}

Estimating the instanton prefactor using the scale of the black hole seed gives this branching ratio
as \textcolor{black}{\cite{Gregory:2023eos}}
\be
\frac{\Gamma_D}{\Gamma_H} \sim 88 M^{3/2} \sqrt{\pi \delta M} e^{-4\pi M\delta M}
\label{approxratio}
\ee
which is sketched in Figure~\ref{fig:branch} and corresponds
well with the figures in~\cite{Burda:2015isa,Burda:2016mou},
\textcolor{black}{which were obtained by numerical evaluation}.

Finally, an alternate description of vacuum decay is given by the HM~\cite{Hawking:1981fz} instanton. This type of decay applies to potentials that have a very flat 
barrier to decay, and~the transition occurs from the local false vacuum minimum to the
top of the potential barrier, from~which the universe can roll to lower true vacua. The~HM
rate is, then, given by $\Gamma_{F\rightarrow T}\sim e^{-B}$, where B is simply the difference 
in the Euclidean actions of the two vacuum de-Sitter spacetimes:
\be
I_{HM} = I_T-I_{F} = S_F- S_T = \frac{\pi}{G} \left [ \ell_F^2 - \ell_T^2 \right]
\ee
which, again, is a Boltzmann factor related to the drop in entropy incurred by the cosmological
horizon area decreasing. The~corresponding process for the black hole Hawking--Moss (BHHM) 
instantons was considered in~\cite{Gregory:2020cvy,Gregory:2020hia}, where the action includes
the black \mbox{hole horizon}:
\be
B_{F\rightarrow T}=I_T-I_F=[S_{CH}+S_{BH}]_F-[S_{CH}+S_{BH}]_T \;.
\ee

By using the reparametrisation:
\begin{equation} \label{horizons}
r_c = \frac{2}{\sqrt{3}}\ell\cos\left(\tfrac{\pi}{3}-b\right), \quad 
r_h = \frac{2}{\sqrt{3}}\ell\cos\left(\tfrac{\pi}{3}+b\right), \quad b 
= \frac{1}{3}\cos^{-1}\left(\frac{3\sqrt{3}GM}{\ell}\right).
\end{equation}
the BHHM transition action can be compactly expressed as
\begin{equation}
B = \pi \left[{\textstyle{\frac{4}{3}} \left(\ell_F^2 - \ell_T^2\right) 
- \frac{2}{3}\ell_F^2 \cos(2b_F) + \frac{2}{3} \ell_T^2\cos(2b_T)}\right]
\end{equation}

So, for~example, as~$M$ runs from $0$ to its Nariai maximum $\ell/3\sqrt{3}$, $b$
drops from $\pi/6$ to $0$ and the action drops from $\pi \ell^2$ to $2\pi \ell^2/3$.
It is, therefore, easy to see that, in tunnelling from a black hole seed to a pure de-Sitter
state, the action drops; therefore, black holes increase the probability of a transition. \textcolor{black}{The full details can be found in~\cite{Gregory:2020hia}.}

\section{Seeded Decay in Einstein Gauss--Bonnet Gravity: Bubble~Nucleation}
\label{sec: Extend on EGB}

While most of the phenomenologically relevant computations for vacuum decay have been 
discussed in four spacetime dimensions, it is also interesting to consider how extra dimensions
can impact vacuum decay. General considerations of how large extra dimensions could
impact black hole seeded vacuum decay were explored in~\cite{Burda:2015isa,Cuspinera:2018woe,Cuspinera:2019jwt} for both noncompact and braneworld-type scenarios in the context of Einstein gravity. However, allowing for the dimensionality
of spacetime to be arbitrary raises the possibility of including higher curvature terms in the action.
\textcolor{black}{Here, we focus on the addition of the GB term~\cite{Lanczos:1938sf, Zumino:1985dp, Zwiebach:1985uq} as it is the lowest-order Lovelock term
~\cite{Lovelock:1971yv}, which still retains second-order equations of motion. 
Further, since the computation of instanton actions frequently involves a background subtraction, we 
require a Birkhoff-type theorem, so that we know that the relevant background is unique. The~Lovelock terms satisfy this condition, whereas generic higher-order terms do not (as well as typically suffering from a lack of well-posedness).}
In 4D, the~GB term is a topological invariant and~does not alter the dynamics of the theory; 
however, it can potentially change the action, hence is of relevance to a tunnelling process,
where the action determines the probability of decay.
\textcolor{black}{We first briefly review the derivation of the equations of motion from~\cite{Gregory:2023kam} before extending the discussion to include dynamical bubbles. 
By expanding in the GB parameter, $\alpha$, we are able to show that the equations of motion can be re-cast in the same form as the Einstein equations of motion, albeit with different details in the coefficients. 
We recap the computation of the action, then discuss a test case example in arbitrary dimensions in detail, discussing the impact of the $\alpha$ term. 
We finally comment on a peculiarity of EGB black holes in 5D, namely the presence of a ``mass gap'' in the black hole family of solutions.}

\subsection{Seeded Bubbles in~Einstein-Gauss--Bonnet}
\label{subsec: instansol}

The gravitational action of Einstein-Gauss--Bonnet (EGB) gravity is
\be
\beal
I = &- \frac1{16\pi G} \int_{\cal M} d^D x \sqrt{g} 
\left [ {\cal R}- 2 \Lambda +\alpha {\cal L}_{GB}\right] \\
\vspace{0.23cm}
 {\cal L}_{GB} =& R^2 - 4 R_{ab}^2 + R_{abcd}^2 
\label{EGBaction}
\eeal
\ee
where $\alpha$ is a dimensionful coupling constant, here taken to be positive. 
As in the pure Einstein case described previously, we look for bubble trajectories in the thin wall 
approximation, where the vacuum transitions completely and instantaneously across the wall, 
as this will capture the impact of the additional GB term on the~process.

\textls[-20]{As before, there is a Birkhoff theorem for EGB gravity in the presence of a brane
~\cite{Wiltshire:1985us,Charmousis:2002rc, Bogdanos:2010zz},} and~the brane junction 
conditions were discussed in~\cite{Davis:2002gn}, although~were only explicitly derived for
$\mathbb{Z}_2$ symmetry across the wall. The~bubble spacetime is again a wall separating
two spherical vacuum solutions with a seed and remnant mass black hole as in~\eqref{bubblegeom},
but with $d{\Omega}^2_{I\!I} \to d{\Omega}^2_{D-2}$ as appropriate for a higher-dimensional
solution and~with the metric potential now~\cite{Boulware:1985wk, Wheeler:1985nh, Wheeler:1985qd}. 
\be\label{metricfunctions}
f_\pm =1+\frac{r^2}{2\tilde{\alpha}}\left(1-\sqrt{1+
\frac{8\tilde{\alpha} \Lambda_\pm}{(D-1)(D-2)}
+\frac{4\tilde{\alpha}\mu_\pm}{r^{D-1}}}\right),
\ee
with $\tilde{\alpha}=(D-3)(D-4)\alpha$ and~$\mu$ related to the mass through
the Myers--Perry formula~\cite{Myers:1986un}:
\be
\mu = \frac{16\pi G M}{(D-2) {\cal A}_{D-2}}\;,
\ee
and ${\cal A}_{D-2}$ being the area of a unit $(D-2)-$sphere.

The action of the bubble is the sum of the parts from the bulk on each side given by \eqref{EGBaction}
together with the wall piece, including both the wall tension and the geometrical Gibbons--Hawking-type terms, which now include additional EGB contributions~\cite{Davis:2002gn, Brihaye:2008ns}:
\be
\beal
I_{brane}=
\int_{\text{brane}} d^{D-1} x \sqrt{h} \left (  \sigma +
\frac{1}{8 \pi G} \left( \Delta K -2\alpha \left[2\mathcal{G}_{ab} \Delta K^{ab}
-\Delta \mathcal{J}\right] \right) \right)
\end{aligned}
\label{Ibrane}
\end{equation}
where ${\cal J}$ is the cubic extrinsic tensor:
\be
{\cal J}_{ab} = \frac13 \left ( 2 K K_{ac} K^c_{\; b} + K_{cd}K^{cd} K_{ab} 
-2K_{ac}K^{cd}K_{db} - K^2 K_{ab} \right).
\ee

\textls[-15]{The generalised Israel junction conditions at the wall are obtained by varying \mbox{this action}:}
\be
\Delta K_{ab} - \Delta K h_{ab} + 2 \alpha \left [3 \Delta{\cal J}_{ab}
- \Delta{\cal J} h_{ab} - 2 {\cal P}_{acbd} \Delta K^{cd} \right] = 8 \pi G \sigma h_{ab}
\label{GBIsrael}
\ee
with 
\be
{\cal P}_{abcd} = {\cal R}_{abcd} + 2  {\cal R}_{b[c} h_{d]a} 
-2  {\cal R}_{a[c} h_{d]b} +  {\cal R} h_{a[c} h_{d]b}
\ee
the divergence-free part of the intrinsic Riemann tensor. The~equations of motion, while simple for a spacetime
reflection symmetric around the wall
now become far more involved when we have the situation of a false--true vacuum and 
the interior--exterior of the bubble. 
Writing the general wall trajectory as $R(\lambda)$
and computing the extrinsic and intrinsic curvatures,~\eqref{GBIsrael} becomes
\be
\beal
-\frac{8\pi G \sigma R}{(D-2)} = - 2 \bsig R =&
\left ( \sqrt{f_+ - {\dot R}^2} -  \sqrt{f_- - {\dot R}^2} \right)
\left [ 1 + \frac{2\tila}{R^2} - \frac{4\tila}{3}\frac{{\dot R}^2}{R^2} \right]\\
&- \frac{2\tila}{3R^2} \left ( f_+  \sqrt{f_+ - {\dot R}^2} - f_-  \sqrt{f_- - {\dot R}^2}\right)
\eeal
\ee 

\textls[-20]{As with the Einstein case, this equation can be manipulated into a Friedmann-like equation for $R$,
which is most compactly written in terms of the variable 
$X=(1-\dot{R}^2)/R^2$ \cite{Gregory:2023kam}:}
\be
\beal
0 &=  \tilde{\alpha}^2 X^3 +  \tilde{\alpha} X^2 \left ( \frac16
- \frac{(S_+^2-S_-^2)^2}{256 \tilde{\alpha} \bar{\sigma}^2} \right)\\
&+ \frac{X}{4} \left ( 3 - \frac{3(S_+^2+S_-^2)}{8}
- \frac{(S_+^2 - S_-^2)(3S_+^2-3S_-^2 + S_+^3 - S_-^3)}
{192 \tilde{\alpha} \bar{\sigma}^2} \right ) \\
&+ \frac{(8 - 3S_+^2-3S_-^2 - S_+^3 - S_-^3)}{64 \tilde{\alpha}} 
- \frac{(3S_+^2-3S_-^2 + S_+^3 - S_-^3)^2}{9216 \tilde{\alpha}^2 \bar{\sigma}^2}
-\frac{9\bar{\sigma}^2}{16}
\eeal
\label{cubicFriedmann}
\ee
where, for~compactness of notation, we introduce
\be
S_\pm = \sqrt{1+\frac{8\tilde{\alpha} \Lambda_\pm}{(D-1)(D-2)}
+\frac{4\tilde{\alpha}\mu_\pm}{r^{D-1}}}
= \sqrt{ 1 + \frac{4\tilde{\alpha}}{r^2} (1-f_{_\pm E})}
\ee
which is related to the standard Einstein
potential $f_E = 1 - \mu/r^{D-3} -2 \Lambda r^2/(D-1)(D-2)$. 

While one can in principle solve the cubic \eqref{cubicFriedmann} to find the solution for $X$ 
(by identifying the root that is connected to the Einstein solution as $\alpha\to0$), 
it proves useful to identify the ${\cal O}(\alpha)$ correction to the trajectories by 
expanding \eqref{cubicFriedmann} in $\tilde \alpha$ to obtain
a Friedmann-like equation for $R$:
\be
\frac{{\dot R}^2}{R^2} = \frac{{\bar f_E}}{R^2} - \frac{(\Delta f_E)^2}{16 R^4  {\bar\sigma}^2}
- {\bar\sigma}^2 + \frac{\tilde \alpha}{3R^4} \left \{  3({\bar f_E}-1)^2 + \frac{(\Delta f_E)^2}{4}
+12  R^2 {\bar\sigma}^2 (1-{\bar f_E}) + 8 R^4 {\bar\sigma}^4
\right\}
\label{frwalpha}
\ee
where ${\bar f}_E = (f_++f_-)/2$.

\textcolor{black}{Note that both this ${\cal O}(\alpha)$ equation, as~well as the original cubic in $X$ are quadratic in $\dot{R}$. This means that, to obtain the real time evolution of the bubble once it has nucleated, we simply analytically continue back to Lorentzian time, $\tau \to -i t$, to~determine the real time evolution of the bubble. A ``round'' Euclidean bubble will correspond to a hyperbolic Lorentzian solution, where the bubble starts to expand upon formation. A ``static'' Euclidean bubble will nucleate to an initially static Lorentzian bubble, but~this is an unstable configuration, so we expect that, on~average, half of these bubbles will, then, expand to complete the phase transition.}

From \textcolor{black}{\eqref{frwalpha}}, we can manipulate the equation of motion for $R$ in a fashion 
similar to~\cite{Gregory:2013hja}, where the effective scale of $R$ was defined
in terms of parameters $\gamma$ and $a$ (the 
 notation of~\cite{Gregory:2013hja}
was $\alpha$, which we have modified here to avoid confusion). To~include 
the GB parameter $\alpha$, define:
\be
\bsig_e^2 = \bsig^2 ( 1-8\tila \bsig^2/3) \;\;,\quad 
\frac{1}{\ell^2_e} = \frac{2\Delta\Lambda(1-4\tila\bsig^2)}{(D-1)(D-2)}\;\;,\quad
\gamma_e = \frac{4\bsig_e \ell^2_e}{1 + 4 \bsig^2_e\ell^2_e}\;\;,
\ee
and
\be
a^2 = 1 + \frac{2 \Lambda_- \gamma_e^2}{(D-1)(D-2)} \left [ 1 - 4 \tila \bsig^2 
-\frac{2\Lambda_+ \tila}{(D-1)(D-2)} \right]
\ee

Now, defining ${\hat R} = aR/\gamma_e$ and ${\hat \lambda} = a\lambda/\gamma_e$, 
the equation of motion takes the form:
\be
\left ( \frac{d {\hat R}}{d{\hat\lambda}} \right)^2 = 1 - \left ( {\hat R} 
+ \frac{\hat C}{{\hat R}^{D-2}}
\right)^2 - \frac{{\hat B} - 2 \hat C}{{\hat R}^{D-3}}
\label{eqgenform}
\ee
where
\be
\beal
\hat{B} &=\left ( \frac{a}{\gamma_e} \right)^{D-3} \left [ {\bar\mu} \left ( 1 
- 4 {\tilde\alpha} \bsig^2- \frac{4 {\bar\Lambda}\tila}{(D-1)(D-2)} \right )
+ \frac{\Delta \mu}{8\bsig^2_e \ell_e^2}\right]
\\
\hat {C} &= \left ( \frac{a}{\gamma_e} \right)^{D-2} \left [
\frac{(\Delta\mu) (1-4 \tila \bsig^2)}{4 \bsig_e} 
- 2\tila \frac{\mu_+\mu_-\bsig_e}{\Delta\mu}\right ]
\eeal
\ee

This format is the same as for the Einstein case, although~the coefficients are corrected by
$\tilde\alpha$; thus, the principle for solving the instanton equations is the same. The~
qualitative picture, of~the preferred instanton being a modified CDL bubble with a black hole seed,
but no remnant for a small mass, and~a static instanton with a remnant black hole for larger masses
remains; however, the~relation between the seed and remnant masses will be modified by $\tilde{\alpha}$. 

\subsection{Bubble~Actions}

While the general cubic equation for $X$ is algebraically involved, we can still derive an 
expression for the action of a general instanton in terms of the wall trajectory $R(\lambda)$,
then specialise to the case where $R(\lambda)=R_0$ is constant. We also expect that 
the ratio $\gamma_e/a$ will still broadly represent the scale of the wall~motion.

To compute the action, the~procedure is the same as for the Einstein case, where we first
compute the action of the background (seed) geometry with the proviso that we compute at
an arbitrary periodicity of Euclidean time, so that we can match to the exterior of the bubble 
spacetime. As~with the Einstein instanton, this shifting of periodicity means we have to 
regularise the resulting conical deficit at the horizon, a~procedure detailed in~\cite{Gregory:2023kam}.
The result is that we can write the seed action as
\be
I_{seed} = I_{c} - S_{BH}(M_+).
\ee
where $S_{BH}$ is the entropy of the black hole
\be
S_{BH} = \frac{{\cal A}_{D-2}r_h^{D-2}}{4 G} 
\left ( 1 + \frac{2(D-2) \tilde{\alpha}}{(D-4)r_h^2} \right) 
\label{rhentropy}
\ee
and $I_c$ represents a ``large $r$'' contribution, which depends on the asymptotic 
structure of the spacetime. While we do not need to know its exact form, as~this term
will be the same for both bubble and seed spacetimes, its general form is:
\be
I_c = \begin{cases}
\frac{\beta {\cal A}_{D-2}}{16\pi G} r_c^{D-4} f'(r_c) 
\left ( r_c^2 + \frac{2(D-2) \tilde{\alpha}} {(D-4)} (1-f(r_c)) \right) & \Lambda \leq~0\\
- S_{cos} & \Lambda>0
\end{cases}
\label{rccontribution}
\ee

For the bubble geometry, the~fact that the bulk integral \eqref{EGBaction} reduces to a total derivative
on shell means that it only contributes an entropy term at the black hole horizon $r_h$, the~same term as
the seed at the large $r$ cutoff $r_c$, together with contributions evaluated on each side of
the wall:
\be
\beal
I_{bulk, {\cal W}} = 
& - \frac{{\cal A}_{D-2}}{16\pi G} \int d\lambda R^{D-4} \Bigl [ f'_+ \dot{t}_+
\left ( R^2 + \frac{2(D-2)\tilde{\alpha} } {(D-4)} (1-f_+) \right) \\
& -f'_- \dot{t}_- \left ( R^2 +  (1-f_-) \right ) \Bigr].
\eeal
\ee

Meanwhile, the~wall integral \eqref{Ibrane} can be manipulated using the trace of the Israel 
equations to give
\be
I_{\cal W}= \frac{{\cal A}_{D-2}}{8\pi G} \int d\lambda \frac{R^{D-2}}{(D-1)}
\left ( \Delta K -\frac{6\tilde{\alpha}}{(D-3)(D-4)} \left[2\mathcal{G}_{ab} \Delta K^{ab}
-\Delta \mathcal{J}\right] \right) .
\ee

Giving the total wall contribution as
\be
\resizebox{0.93\hsize}{!}{$
I_{wall} = \frac{{\cal A}_{D-2}}{8\pi G} \int d\lambda R^{D-2} \Delta \Biggl [
-\frac{4(D-2)\tilde{\alpha}}{(D-4)} \frac{\ddot{R}}{R} K_1
+ \left ( \frac{\dot{R}^2}{f} K_0 - \dot{t} \ddot{R}\right)
\left ( 1 + \frac{2(D-2)\tilde{\alpha}(1-f)}{(D-4)R^2} \right) \Biggr ].
$}\ee

Pulling together all these formulae, the~instanton action is computed as
\be\label{bubbleaction}
I_{\cal B} = I_{bubble} - I_{seed}
= S_{BH}(M_+) - S_{BH}(M_-) + I_{wall}.
\ee

In other words, as~with Einstein gravity, the~instanton action consists of an entropy differential
together with a dynamical wall integral, which was shown in~\cite{Gregory:2023kam} to vanish for the static
instanton. Thus, as~for Einstein gravity, the~dominant decay process for most black holes is the static instanton and~has a probability of
\be\label{differinentro}
\mathcal{P}\sim exp[-({S_{seed}}-S_{remnant})/\hbar]
\ee

In 4D, although~the GB term has a vanishing contribution to the gravitational dynamics, it does
change the action of a solution due to the modified expression for entropy:
\be
S_{BH} = \frac{\pi r_+^2}{G} \left ( 1 + \frac{4\alpha}{r_+^2} \right)
=  \frac{\pi r_+^2}{G} +  \frac{4\pi \alpha}{G}
\ee

This constant shift is not relevant for the thermodynamics of the black hole~\cite{Clunan:2004tb}, 
and if tunnelling from a seed to a remnant black hole, these terms will cancel; however,
if we have a topology-changing process such as tunnelling from a seed to no remnant, then
this term will be relevant. Thus, since it is possible to show that a bubble cannot nucleate with 
a black hole inside from a pure false vacuum
\textcolor{black}{\cite{Gregory:2023kam}}, this term always suppresses topology-changing~transitions.

In higher dimensions, the~GB parameter will modify the wall trajectory, as~well as the action, 
so in order to extract the impact of the GB correction, we will explore how the critical static bubble
is impacted by $\alpha$. The~critical static bubble is the instanton with the lowest action for a given
wall tension $\sigma$. It connects the modified CDL branch, which represents the preferred instanton
at very low seed masses, where the interior of the bubble has no black hole, to the static branch, where
there is both a seed and remnant black hole; the critical instanton is, therefore, static, but~has no
remnant, and~the seed mass is fully determined by the vacuum energies and wall~tension.

For simplicity, we will explore tunnelling from a positive vacuum energy with the seed black
hole to the Minkowski vacuum. Therefore, $\Delta \mu = \mu_+ = 2 \bar\mu$, and~
$\Delta \Lambda = \Lambda_+= 2\bar\Lambda$. Thus, $a=1$, and~the parameters
in \eqref{eqgenform} simplify to
\be
\hat{B} = \frac{\mu_+(1-4 \tila \bsig^2)}{2 \bsig_e\gamma^{D-2}_e} = 2 \hat {C}
\ee

Thus, the static instanton constraints read
\be
\beal
1 &= \left ( {\hat R} + \frac{\hat C}{{\hat R}^{D-2}} \right)^2 \\
0&= 2 \left ( {\hat R} + \frac{\hat C}{{\hat R}^{D-2}} \right)\left (1- (D-2) \frac{\hat C}{{\hat R}^{D-1}} \right)
\eeal
\ee
identical in form to the Einstein case. These are solved by
\be
{\hat C} = \frac{(D-2)^{D-2}}{(D-1)^{D-1}}\quad ; \qquad
{\hat R} = \frac{(D-2)}{(D-1)} 
\ee

Hence,
\be
\mu_{crit}= \frac{4 \bsig_e \gamma_e^{D-2}}{(1 - 4 \tila\bsig^2)}\frac{(D-2)^{D-2}}{(D-1)^{D-1}}
\ee
from which one can see that the critical instanton seed mass increases with $\tilde{\alpha}$ by expanding
the expressions for the effective parameters $\bsig_e,\gamma_e$:
\be\label{massvar}
\frac{\delta \mu_{crit}}{\mu_0} = \frac83
\tila \bsig^2 \left [ 1 + (D-2)(1 - 2 \bsig \gamma_0)
\right ]
\ee
where the subscript ``$0$'' indicates the values of the various functions
at $\tila=0$ and~$\mu_0$ is the critical mass at $\tila=0$:
\be
\mu_0 = \frac{(4\bsig\ell_0)^{D-1}\ell_0^{D-3}}{(1+4\bsig^2\ell_0^2)^{D-2}} \frac{(D-2)^{D-2}}
{(D-1)^{D-1}} 
\ee

However, the~instanton action is determined by the entropy of the seed mass, which, in turn, is determined by the horizon radius, which actually \emph{decreases} as $\tila$ is switched on. 
Thus, the actual incremental change in the action is
a combination of these effects and~further complicated by the nature of the polynomial determining the horizon in the various dimensions.
Computing the incremental change in the horizon radius and critical mass, we arrive at an expression for the entropy shift at small $\tila\lesssim \mu_0^{2/(D-3)}$:
\be \label{entropychange}
\frac{\delta S}{S_0} = \frac{(D-2) \tila}{r_0^3 f'_E(r_0)} \left ( \frac{2r_0f'_E(r_0)}{(D-4)}  
+ \frac{8\mu_0\bsig^2}{3r_0^{D-5}}
\left [ 1 + (D-2)(1 - 2 \bsig \gamma_0)
\right ] - 1 \right)
\ee

\begin{figure}[H]

\begin{adjustwidth}{-\extralength}{0cm}
\centering 
\includegraphics[height=2.2in, width=3in]{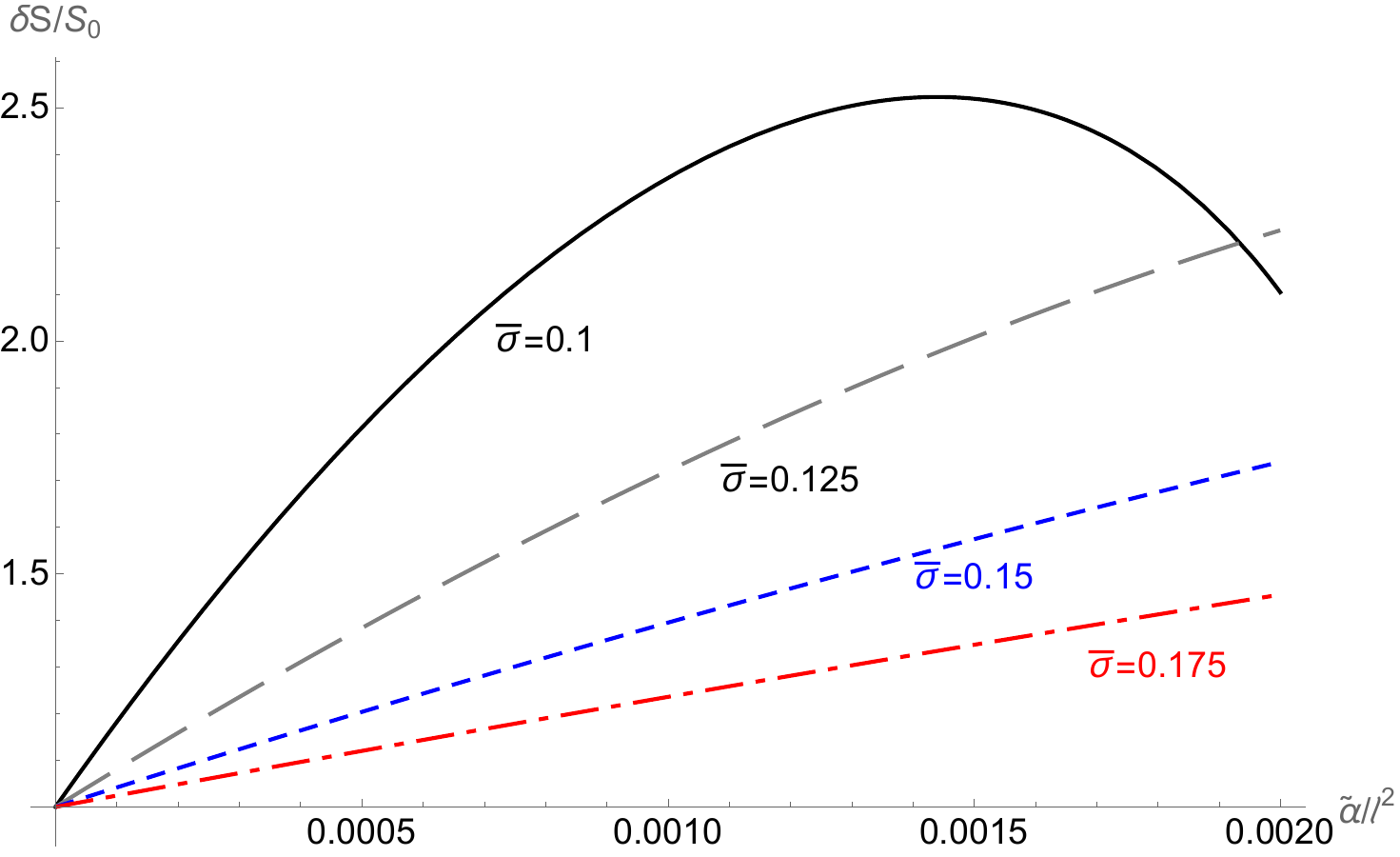}
\includegraphics[height=2.2in, width=3in]{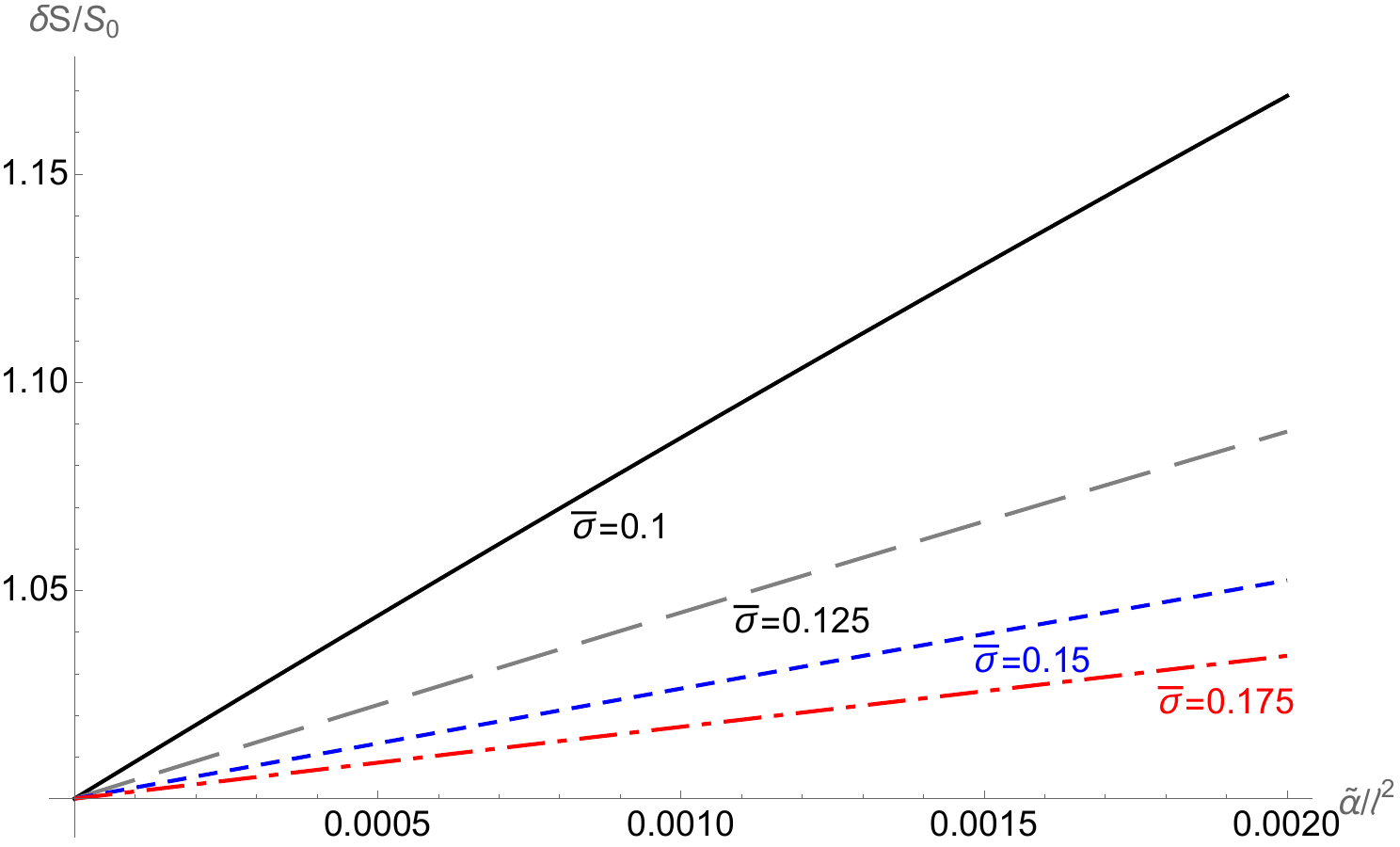}
\end{adjustwidth}
\caption{The 
 shift in entropy of the critical instanton in 5 (\textbf{left}) and 6 (\textbf{right}) dimensions as a function of $\tila/\ell^2$ 
for the given bubble wall tensions indicated, $\bar\sigma=0.1, 0.125, 0.15, 0.175$, respectively. } 
\label{fig:entropyvar} 
\end{figure}

The shift in entropy is proportional to $\tila$, as expected; the other key dependence is on $\bsig\ell_0$, which, in turn, feeds into $\mu_0$.
As discussed in~\cite{Burda:2015yfa}, $2\bsig\gamma_0<1$; hence, the critical mass becomes very strongly damped for low $\bsig\ell$ at large $D$; hence, the range of $\tila$ for which \eqref{entropychange} is relevant becomes small. This is reflected in the range for $\tila\ell_0^2$ in Figure~\ref{fig:entropyvar}.

Exploring the entropy change \eqref{entropychange} for various $D$ shows that the shift
in entropy with $\tila$ is positive. Thus, the~GB term lowers the tunnelling amplitude. The~picture for $5D$ is more nuanced, however. In~5D, the~horizon radius can be simply found as 
\be
r_h= \frac{\ell_0}{\sqrt{2}}
\left(1-\sqrt{1 + \frac{4(\tila-\mu)}{\ell_0^2}}
\right)^{1/2}
\label{5Dradius}
\ee

Clearly, $r_h\to0$ as $\mu\to\tila$, i.e.,  at finite values of $\mu_{crit}$; hence, $\bsig$, and~there is
no horizon for $\mu<\tila$. Thus, we have an interesting
``mass gap'' in the 5DGB black hole family.
Spacetimes with $\mu<\tila$, i.e.,  ADM mass
below $3\pi \alpha/4G$, do not have a horizon, but~instead have a solid angle deficit as $r\to0$.
This phenomenon has been noticed in the study of gravitational collapse in EGB gravity~\cite{Deppe:2012wk, Deppe:2014oua, Frolov:2015bta}. 
This makes 5D crucially different from higher dimensions, in~that, while the 
entropy of the seed does initially increase with $\tila$, for~low $\bsig\ell_0$, we see
that the entropy function (calculated exactly, rather than using \eqref{entropychange}) turns over 
and decreases, eventually dropping to zero as this mass limit is hit. 
We, therefore, expect that the semi-classical approximation should not be valid for small seed black~holes.

In Figure~\ref{fig:entropyvar}, we show the 
variation in entropy of the seed black hole (which directly translates into the instanton action)
as $\tila$ is switched on for both 5 and 6 dimensions. Although~the generic picture is that 
the entropy increases as $\tila$ increases, the~one exception is for $D=5$ and smaller $\bsig$. 
As discussed above, caution should be used in pushing to too low a $\bsig$.

\unskip

\section{Seeded Decay in Einstein Gauss--Bonnet Gravity: Hawking--Moss}
\label{sec:HM instanton}

\textcolor{black}{In this section, we briefly review Hawking--Moss tunnelling with a GB term~\cite{Gregory:2023kam}, expanding on the distinctiveness of the 5D results.}
The HM instanton is a transition from a false vacuum to a higher vacuum energy, 
from which the universe can roll to lower true vacua. While it is not possible to tunnel from no
black hole to a black hole for bubbles, this can happen for BHHM~transitions.

The HM and BHHM transition probabilities are easy to compute, as~one is comparing the actions for 
Schwarzschild GB de-Sitter geometries, where the cosmological constants are determined by
the vacuum energy in the false vacuum (denoted by the subscript $F$) and the vacuum energy at
the local maximum of the potential (denoted $T$) that the universe instantaneously jumps to 
in this transition. The~action of a Schwarzschild GB dS geometry was shown in the previous
section to be 
\be
I_{SGBDS} = - S_{CH} - S_{BH}
\ee
i.e.,  minus the sum of the entropies of the cosmological and black hole horizons; thus,
the tunnelling rate, dominated by the Boltzmann factor, $\Gamma_{F\rightarrow T}\sim e^{-B}$,
is given by 
\textcolor{black}{
\be
\label{BHHM probablity}
\beal
B_{F\rightarrow T}
& =I_T-I_F=[S_{CH}+S_{BH}]_F-[S_{CH}+S_{BH}]_T \\
&= \frac{{\cal A}_{D-2}}{4G} 
\Bigl [ r_{CF}^{(D-2)} + r_{BF}^{(D-2)} 
- r_{CT}^{(D-2)} - r_{BT}^{(D-2)} \\
&\;\;\; + \frac{2(D-2) \tilde{\alpha}}{(D-4)} 
\left ( r_{CF}^{(D-4)} + r_{BF}^{(D-4)} 
- r_{CT}^{(D-4)}- r_{BT}^{(D-4)} \right) \Bigr]
\eeal\ee
where (for example) $r_{CT}$ stands for the cosmological horizon radius of the nucleated universe at the top of the potential and~$r_{BT}$ for the black hole horizon radius in the \mbox{seed~universe}.}

\textcolor{black}{
In general, in 4D, the~GB term does not affect the BHHM transition;} however, it will 
suppress tunnelling from a black hole seed to a pure GB de-Sitter state. This gives a lack of 
continuity in the most-likely instanton, where tunnelling to an infinitesimally small mass black
hole is preferred over tunnelling to the pure vacuum, although~one should not apply this
semi-classical approach for black holes around the Planck scale.
\textcolor{black}{On the other hand, note
that, now, tunnelling from a pure de-Sitter false vacuum to a Schwarzschild--de-Sitter (SdS) spacetime at the top of the potential is allowed as an HM transition. For~these 
(topology-changing) configurations, the action is now lowered by $\delta I \sim 4\alpha/G - 
2\pi \ell M$ relative to the pure (black-hole-free) HM transition; thus, again, there is a preferred transition to 
a vanishingly small mass black hole universe.
While one might imagine that this small black hole would rapidly evaporate, there is a conflict in constructing a semi-classical solution with a Planckian-scale black hole.}

\begin{figure}[H]
\includegraphics[height=3in, width=3.5in]{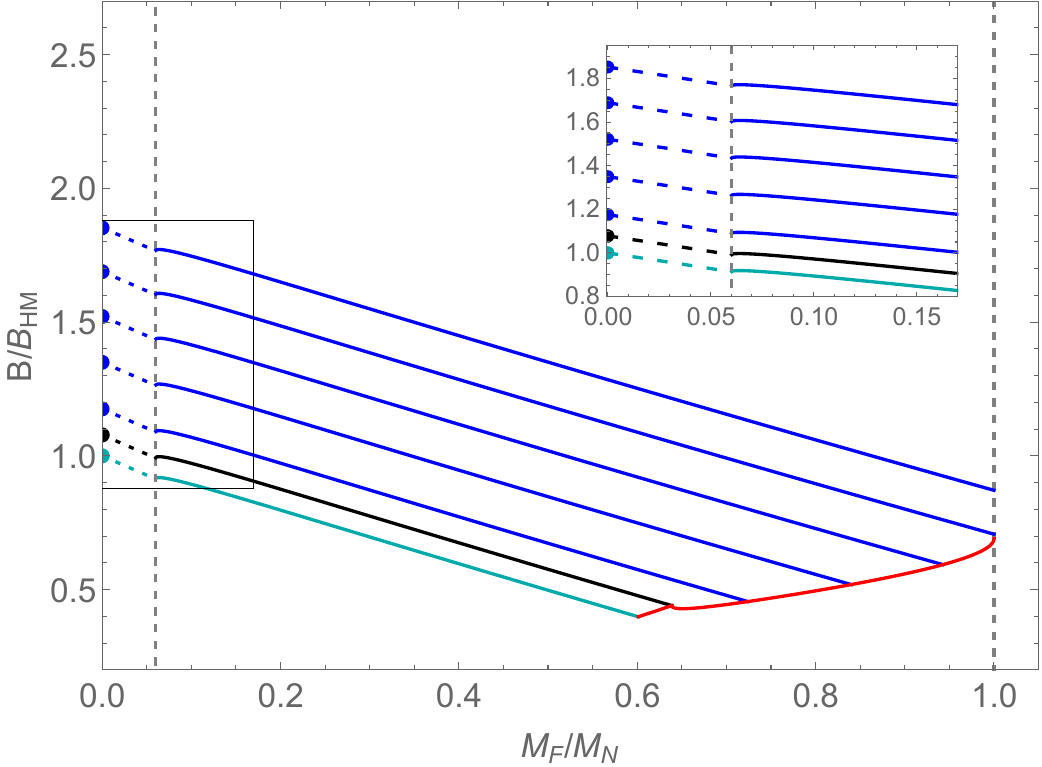}
\caption{$B/B_{HM}$ as 
 a function of seed mass in 
5D for parameter values $\ell_F=5$, $\ell_T=4.5$, and~$\alpha=0.2$, where the 
mass gap is $M_C\approx0.06M_N$. 
The black line indicates $M_{C}$. (Figure 2
from~\cite{Gregory:2023kam}).}
\label{fig:HMD5}
\end{figure}

\begin{figure}[H]
\includegraphics[height=3in, width=3.5in]{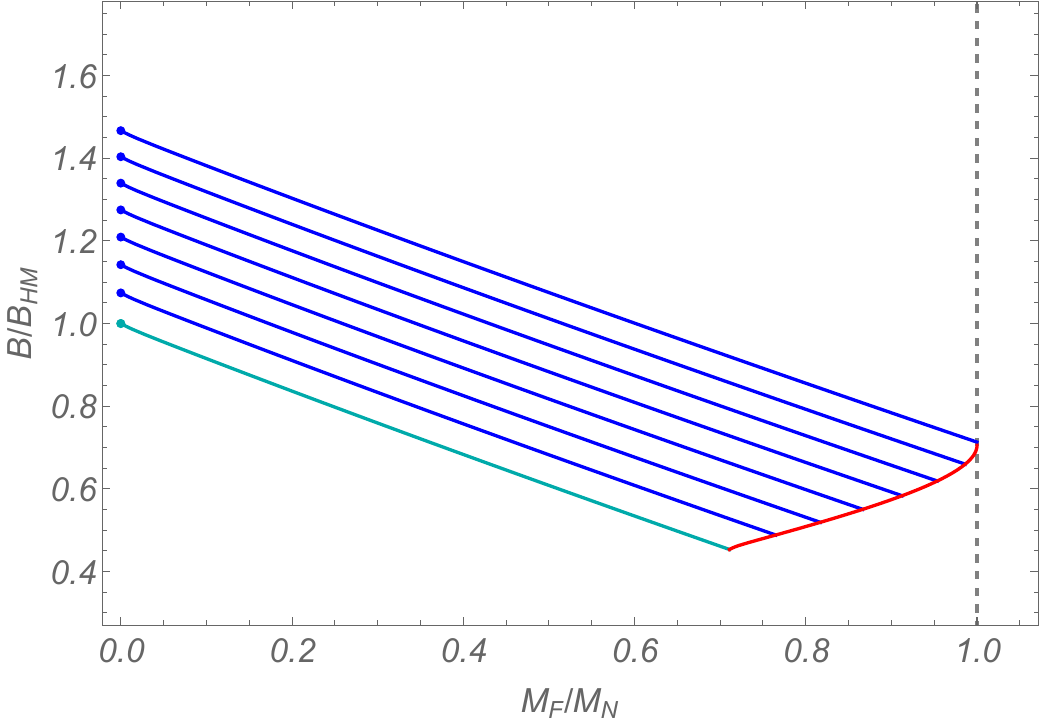}
\caption{$B/B_{HM}$ as a function of seed mass in 6D
shown for parameter values $\ell_F=5,\ell_T=4.5$ and $\alpha=0.2$ (Figure 3 from~\cite{Gregory:2023kam}).
}
\label{fig:HMD6}
\end{figure}

In higher dimensions, the~impact of seeded decay in BHHM transitions can be computed \textcolor{black}{straightforwardly from \eqref{BHHM probablity} (once the horizon radii are found),} 
and the results are shown in Figures~\ref{fig:HMD5} and \ref{fig:HMD6} for 5D and 6D, respectively.
An interesting feature appears in 5D due to the ``mass gap'' for black hole solutions.
As already noted, in~$D=5$, there is a lower limit to the mass a \emph{black hole} can have, 
as the horizon radius $r_h\to0$ as $M \to 3\pi \alpha/4G$. 
Below this mass, the~solution does not have a horizon, but,~instead, contains a solid angle deficit as $r\to0$.
This causes the dependence of the HM tunnelling amplitude with seed mass to have a discontinuity in
its derivative (though it is doubtful this is physically relevant). 

\textcolor{black}{Looking in a bit more detail, the~black hole horizon radius is given by \eqref{5Dradius}, where $\tila = 2\alpha$ and $\mu = 8GM/3\pi$.
Thus, as the critical mass $M_C = 3\pi \alpha/4G$ is approached, the~black hole horizon radius vanishes as
\be
r_b \sim \sqrt{\frac{2\alpha}{M_c}}
\sqrt{ M - M_C}
\ee
giving rise to the kink in the plot. Since there is no black hole 
horizon for spacetimes with mass less than this critical value, there is no entropy contribution from the solid angle deficit at the origin. The~only way the mass contributes to a varying action is through its impact on the cosmological horizon radius.}

As with the Einstein case, the~BHHM action increases with the remnant mass, indicating a larger 
remnant black hole slows down the transition, and~including a black hole seed lowers the action,
indicating a catalysis of the HM process. Figures~\ref{fig:HMD5} and \ref{fig:HMD6} show the
modified BHHM instanton actions in 5D and 6D, respectively. They are qualitatively the same
as for the Einstein case~\cite{Gregory:2020hia}, except~for the mass gap in 5D. The~6D plot is representative of the situation for general higher dimensions.
In both plots, solid blue lines indicate a transition to a given remnant mass and~the cyan line to
no remnant mass. The~red line corresponds to the cosmologically equal area limit and~the dots to 
tunnelling from a pure dS false~vacuum.

\textcolor{black}{The Hawking--Moss instanton is in a very different category to the first-order phase transition of bubble nucleation. 
The formation of bubbles, while a quantum process, is readily understandable as a phase transition through analogy with common physical processes, such as boiling water. 
The Hawking--Moss transition, however, is best understood with a quantum cosmology interpretation. 
Instantaneously, the~whole universe fluctuates into what intuitively seems to be a higher energy state (though, see the discussion around the thermodynamic free energy in~\cite{Gregory:2020hia}). 
Therefore, there is no immediate Lorentzian continuation of the process, which is best viewed as a Boltzmann-suppressed jump. 
Instead, one imagines that, once the universe has taken this jump, it simply continues to roll down the far side of the potential in a classical Lorentzian evolution, akin to an inflating universe.}

\section{Conclusions}

In this paper, we reviewed and extended work on seeded tunnelling in Einstein and EGB gravity. 
From the Euclidean saddle point approximation, black holes considerably enhance the probability of vacuum decay, and~we showed that these results are robust to the addition of higher-order \textcolor{black}{Gauss--Bonnet} gravitational terms in the action, as well as extra dimensions. 
\textcolor{black}{Interestingly, a~more-generic higher-order term in the gravitational action would be problematic from this Euclidean saddle point approach. 
The saddle point instanton computation requires a background subtraction, i.e.,  an evaluation of a ``background'' geometry with the same asymptotics as the instanton. We rely on the uniqueness of the Schwarzschild geometry to perform this subtraction; this arises due to a Birkhoff theorem for spherically symmetric backgrounds. In~the absence of such a theorem, we no longer have the uniqueness of this background
~\cite{Oliva:2011xu,Kehagias:2015ata}. Therefore, it might be that the simplest solution in a particular higher-order theory is not, in~fact, the~lowest Euclidean action solution that one should be using for background subtraction.}

The study of bubble nucleation in EGB gravity was presented using the thin wall approximation, as~the aim was to investigate the gravitational aspects of the problem.
The equations of motion obtained by using the thin wall approximation are a good approximation when the potential barrier is sharp and transitions are rapid and complete.  
However, Higgs vacuum decay is, in fact, an extended thick wall region in which the Higgs deviates from its false vacuum SM value; thus, the Boltzmann suppressed static branch (where the universal result \eqref{differinentro} applies) will be the dominant saddle point decay.
Given the apparently thermal nature of this instanton, a~valid concern is that, if the tunnelling is taking place in a region around a primordial black hole, then we should be taking into account the thermal corrections to the Higgs~potential.

How to precisely compute the thermal effect, however, is a difficult question, as it requires going beyond the saddle point and including not only field theory corrections, but~also considering the gravitational impact. 
There have been several studies on this problem~\cite{Kohri:2017ybt,Hayashi:2020ocn,Shkerin:2021zbf, Shkerin:2021rhy, Strumia:2022jil, Briaud:2022few}, 
exploring whether the thermal fluctuations stabilise the Higgs potential and, thus, preclude decay. 
Qualitative arguments might indicate that tunnelling would be suppressed or removed, 
but these do not consider potentially critical pre-factors 
and do not address the gravitational aspects of the problem in detail. 
Alternately, studies using a Euclidean approach or in lower dimensions leave a more-mixed picture. 
One problem is that the thermal corrections to the Higgs potential are only part of the question; 
one has to consider the cosmological setting more holistically to obtain a more-physically complete picture. 
In~\cite{Hamaide:2023ayu}, the~decay was studied allowing for an interstellar medium, 
which will be impacted by the radiation from an evaporating black hole. 
A hot spot forms, surrounding the primordial black hole, and~provides a constant 
(on the evaporation/tunnelling timescale) temperature higher than the universe, but lower 
than the Hawking temperature of the black hole.
This gives rise to thermal corrections, which tend to  suppress decay while
the conical deficits enhance the decay rate. The~realistic environment, as well as the gravitational 
impacts from black holes suggest that the suppression from thermal corrections is not~complete.

Moving to the higher dimensions and higher curvature gravity considered here, interestingly, the GB term lowers the temperature of a black hole of a given mass relative to its Einstein value. 
This would further mitigate any thermal corrections to the Higgs potential, hence lessening the relevance of such corrections.
We also note that, if there are any couplings of the Higgs to the GB term (analogous to the conformal couplings such as considered in~\cite{Rajantie:2016hkj}), then even in 4D, the GB will now acquire some dynamical input into the different decay channels; this is an interesting and open~question.

\vspace{6pt} 



\authorcontributions{Conceptualization, R.G. and S.-Q.H.; methodology, R.G. and S.-Q.H.; formal analysis, R.G. and S.-Q.H.; investigation, R.G. and S.-Q.H.; writing---original draft preparation, S.-Q.H.; writing---review and editing, R.G.;  supervision, R.G.; All the authors have substantially contributed
to the present work. All authors have read and agreed to the published version of the manuscript.
}

\funding{This work was supported in part by the STFC Consolidated grant ST/P000371/1 and the STFC Quantum Technology grant ST/005858/1 (RG), the~Chinese Scholarship Council and King's College London (SH), the~Aspen Center for Physics, supported by National Science Foundation grant PHY-2210452 (RG),
and the Perimeter Institute for Theoretical Physics (RG). Research at Perimeter Institute is supported by the Government of Canada through the Department of Innovation, Science and Economic Development Canada and by the Province of Ontario through the Ministry of Colleges and Universities.

} 

\dataavailability{Data are contained within the article. 
} 

\acknowledgments{We thank Lewis Croney for useful discussions. 
}

\conflictsofinterest{The authors declare no conflicts of interest. 
} 



\abbreviations{Abbreviations}{
The following abbreviations are used in this manuscript:\\

\noindent 
\begin{tabular}{@{}ll}
GR & General Relativity\\
GB & Gauss--Bonnet\\
SM & Standard Model \\
HM & Hawking--Moss\\
dS & de-Sitter \\
CDL & Coleman--De Luccia\\
BHHM & black hole Hawking--Moss \\

\end{tabular}

\noindent 
\begin{tabular}{@{}ll}
EGB &~~~Einstein-Gauss--Bonnet\\
ADM &~~~Arnowitt--Deser--Misner \\
SdS &~~~Schwarzschild--de-Sitter 
\end{tabular}
}


\begin{adjustwidth}{-\extralength}{0cm}

\reftitle{References}

\PublishersNote{}
\end{adjustwidth}
\end{document}